\documentclass[a4paper]{article}

\usepackage{graphicx}
\usepackage{mathrsfs}  
\usepackage{amsmath}
\usepackage{float}
\usepackage{authblk}
\usepackage{ntheorem}

\newtheorem{hyp}{Hypothesis}

\providecommand{\keywords}[1]
{
	\small	
  \textbf{\textit{Keywords:}} #1
} 

\begin{document}

\title{Towards a full--scale version of Yakhot's model of strong turbulence}

\author[]{Christoph Renner}
\affil[]{Independent Researcher}
\affil[]{\textit {renner\_christoph@web.de}}

\date{15 April 2026} 
\maketitle

\begin{abstract}
We present first elements of an extension of Yakhot's model of strong turbulence \cite{Yakhot:1998:pdf} towards small scales. The analysis is based on an empirically observed relation for even order structure functions which extends from the inertial into the dissipation range. With this relation and Kolmogorov's four--fifth law, models for structure functions of orders two and three can be derived that replicate expected small scale limits and describe the transition from dissipative to inertial range scaling regimes correctly. An additional length scale parameter is introduced by the extension. It marks the crossover point from the inertial to the dissipation range and can be expressed as a function of the Reynolds number. In combination with a recently proposed large--scale extension of Yakhot's model, we ultimately obtain full--scale models for structure functions of second and third order. These expressions are closed--form, do not contain free parameters and are in good agreement with experimental data from the smallest dissipative scales up to the system scale.

\end{abstract}


\section{Introduction}

While substantial contributions have been made in past decades, turbulent fluid motion remains a notoriously challenging subject of research \cite{Sreenivasan:2025:WhenIsItSolved}. A setting often considered in that context is the idealized case of fully developed, homogeneous, isotropic and stationary turbulence. It is usually studied by means of velocity increments, the difference of the velocities at two points in space separated by a certain distance. With $w(x)$ denoting the component of the velocity field $\bf w$ in direction of the separation vector $\bf l$, the longitudinal velocity increment $v(l)$ is defined as $v(l) = w(x+l) - w(x)$. 

The statistical properties of these increments are usually investigated my means of their moments, the so--called structure functions $\mathscr{S}_n$:
\begin{equation}
	\mathscr{S}_n(l) \; = \; \left< \, v(l)^n \, \right>.
\end{equation}

Most models of turbulence assume that in the inertial range, that is for length scales $l$ much smaller than the flow configuration's system scale $L$ and much larger than the scales at which dissipation effects dominate, structure functions follow power laws in $l$:
\begin{equation}
	\mathscr{S}_n(l) \; \propto \; l^{\zeta_n}. \label{scaling}
\end{equation}

In 1941, Kolmogorov \cite{K41} proposed a first model for the scaling exponents $\zeta_n$, the famous linear relation $\zeta_n = \frac{n}{3}$. In the same year, he was able to show that this relation is exact for $n=3$, a result known as Kolmogorov's four--fifth law \cite{Kolmogorov:FourFifth}. For scales $l$ much smaller than the system length $L$ it links the third order structure function with the mean rate of energy dissipation and the spatial derivative of the second order structure function (this term becomes relevant for small scales only):
\begin{equation}
	\mathscr{S}_3(l) \; = \; - \, \frac{4}{5} \, \mathcal{E} \, l \; + \; 6\, \nu \ \frac{\partial}{\partial l} \, \mathscr{S}_2(l). \label{fourfifthWithDimensions}
\end{equation}

Refined models for the scaling exponents have since been developed which, amongst others, predict relations of second \cite{K62} and third \cite{Lvov:2000:AnomalousExponents} order in $n$. A milestone in this development is B. Castaing's closed--form expression for the $\zeta_n$ derived in 1996 from an analogy with thermodynamics \cite{Castaing:1996:Temperature}. His expression is equivalent to a result by V. Yakhot derived two years later \cite{Yakhot:1998:pdf} and can be shown to generalize several of the most relevant theories of turbulence \cite{Me:2012:GeneralScaling}.

The model by Yakhot exhibits an additional interesting feature: It comprises a term that describes the influence of the flow's large scale properties on the inertial range. Important aspects of the transition from small to large scales, such as in particular the decline of odd--order structure functions to zero, are correctly described by the model. On the other hand, the convergence of even order structure functions towards their constant large scale levels is not captured. 

Recently \cite{Me:2025:LargeScaleYakhot}, a phenomenological extension of the model was proposed that closes this gap. Two additional parameters were introduced and determined from large--scale boundary conditions. The extended model describes the transition from inertial to large scales for structure functions of any order correctly and links the prefactors of the scaling laws (\ref{scaling}) to the rms of large--scale velocity fluctuations.	

Both the original and the extended Yakhot model do not cover dissipative effects. These become dominant for small scales and lead to scaling laws that are distinctly different from those in the inertial range. In this article a first step towards an extension of the model framework to dissipative scales is presented. Based on an analysis of experimental data, a relation between spatial derivatives of even order structure functions and the next higher odd order structure function is found which extends from the dissipation through to the inertial range. 

The system of equations is under--determined in general, but Kolmogorov's four-fifth law provides a closure for the equations of order two and three. The expressions for the structure functions describe experimental data correctly over the entire range of length scales. The model extension introduces a characteristic length scale which is found to be a function of the Reynolds number and can be interpreted as the transition point from dissipative to inertial scaling regimes.

The paper is organized as follows: After a brief recap of the large--scale extension of Yakhot's model in section \ref{largeScaleModel}, a summary of the general constraints and model independent small--scale limits is given. Section \ref{evenSmallScaleTerm} then focuses on the key experimental result that forms the basis of all further analyses. Conclusions for the second order structure function in the limit of small scales are given next, before full-scale models for the structure functions of order two and three are derived in section \ref{fullScaleModel}. A summary and discussion of results conclude the article.

Experimental results presented in this paper have been derived from a data set measured in a cryogenic axisymmetric helium gas jet. This set has also been used for the recent study \cite{Me:2025:LargeScaleYakhot} on the large--scale extension of Yakhot's model, further details on the experimental setup and the data can be found in \cite{Chanal:2000:HeliumJet, Me:2002:Universality}.

\section{Large--scale extension of Yakhot's model}\label{largeScaleModel}
Following the conventions introduced in the recent paper on a phenomenological large--scale extension of Yakhot's model \cite{Me:2025:LargeScaleYakhot}, all equations will in the following be expressed in terms of the dimensionless quantities $r$, $u$ and $S_n$ defined as:
\begin{equation}
	r = l/L, \qquad  u = v/\sigma, \qquad S_n(r) = \frac{\mathscr{S}_n(l/L)}{\sigma^n}.
	\label{uAndR}
\end{equation}
Here, $\sigma$ denotes the rms of the velocity fluctuations $w$ and $L$ is the largest relevant scale of the system\footnote{
	This scale, the so--called system scale, is not identical with the often used integral length scale which is also usually denoted by $L$. The integral scale marks the upper end of the inertial range and is significantly smaller than the system scale as used in this paper. See \cite{Me:2025:LargeScaleYakhot} for details on how the system scale is defined and determined from experimental data.
} 
at which velocity fluctuations have decorrelated and structure functions have converged towards their large--scale limit values.

In order to capture the large--scale convergence of structure functions, two additional terms $D$ and $C$ were introduced in \cite{Me:2025:LargeScaleYakhot} into Yakhot's original equation for the probability density function $p(u,r)$ of the velocity increment. The extended model equation reads
\begin{equation}
	B \frac{\partial p}{\partial r} \, - \, \frac{\partial}{\partial u} 
	\left\{ u \frac{\partial p}{\partial r} \right\} \; = \; - \frac{A}{r} 
	\frac{\partial}{\partial u} \Big\{ u p \Big\} \, + \, 
	\frac{\partial^2}{\partial u^2} \Big\{ \, \big( \, u \, c(r) + D \, \big) \, p  \, \Big\}.
	\label{largeScaleExtenssion}
\end{equation}
where the function $c(r)$ is defined as
\begin{equation}
	c(r) \; = \; r \, (1-r) \, C \label{cDefinition}
\end{equation}
with $C$, as well as $A$, $B$ and $D$, being constants\footnote{Yakhot's original model can be recovered by setting $D=0$ and $c(r) = 1$}. 

With the help of Kolmogorov's four--fifth law and large--scale boundary conditions, in particular the condition of vanishing slopes at the system scale, these parameters can be expressed in terms of only one independent parameter. While the choice is arbitrary in principle, $A$ yields the clearest formulation:
\begin{eqnarray}	
	B \; & = & \; 3 \, (A-1) \nonumber \\
	C \; & = & \;  \frac{1}{5} \, A \, \epsilon \, \frac{1-\zeta_2}{F(1)} \; \approx \; \frac{3}{5} \, A \, \epsilon  \nonumber \\
	D \; & = & \, - 2 \, A. \label{modelParameterization}
\end{eqnarray}
Here $\epsilon = \mathcal{E} \frac{L}{\sigma^3}$ is the dimensionless mean rate of energy dissipation, which for the data set considered here has a value of $\epsilon \approx 2.3$ \cite{Me:2025:LargeScaleYakhot}. The function $F(r)$ is defined in eq. (\ref{fOfRdef}).

By multiplication of eq. (\ref{largeScaleExtenssion}) with $u^n$ and integration with respect to $u$, the equation for the structure function $S_n(r)$ of order $n$ is obtained as
\begin{eqnarray}
	\frac{\partial}{ \partial r} S_{n}(r) \; & = & \; \frac{\zeta_{n}}{r} \, S_{n}(r) \, + \, z_{n} \, c(r) \, S_{n-1}(r) \, + \, z_{n} \, D \, S_{n-2}(r), \nonumber \\
		& = & \; \frac{\zeta_{n}}{r} \, S_{n}(r) \, + \, z_{n} \, c(r) \, S_{n-1}(r) \, - \, 2(n-1) \,\zeta_n \, S_{n-2}(r), \label{largeScaleExtendedSun}
\end{eqnarray}
with
\begin{eqnarray}	
	\zeta_{n} \;  & =  & \; \frac{A \, n}{3 \, (A-1) \, + \, n} \; \underset{n \rightarrow \infty}{=} \; A, \nonumber \\ 
	z_{n} \;  & =  & \; \frac{n \, (n-1) }{3 \, (A-1) \, + \, n} \; = \; \frac{n-1}{A} \, \zeta_{n}. \label{zetaAndZn}
\end{eqnarray}
The $\zeta_{n}$ are the aforementioned Castaing exponents that generalize three of the most important theories of turbulence \cite{Me:2012:GeneralScaling}. For second and third order the solutions of eq. (\ref{largeScaleExtendedSun}) with boundary conditions $S_2(1)=2$ and $S_3(1)=0$ are:
\begin{eqnarray}
	S_{2}(r) \;  & = &  \; \frac{2}{1-\zeta_2} \, \left( \, r^{\zeta_2} \, - \, \zeta_2 \, r \, \right), \label{ExtS2Solution} \\
	S_{3}(r) \;  & = &  \; - \, \frac{4}{5} \, \epsilon \, r \,  \left\{ \, 1 - \, \frac{F(r)}{F(1)} \right\}, \label{ExtS3Solution}
\end{eqnarray}
with
\begin{eqnarray}
	F(r) \;  & = & \; \frac{1}{1+\zeta_2}  \, r^{1+\zeta_2} \, - \, \frac{\zeta_2}{2} \, r^2 \, - \, \frac{1}{2+\zeta_2} \, r^{2+\zeta_2} \, + \, \frac{\zeta_2}{3} \, r^3. \label{fOfRdef}
\end{eqnarray}

To complete the model specification, the independent parameter $A$ was set to $A=7.3$, in accordance with \cite{Sreenivasan:2021:DynamicsFromNsEq} where this value was found to be the limit of the scaling exponents $\zeta_n$ for $n \rightarrow \infty$ (c.f. eq. (\ref{zetaAndZn})).

\section{General constraints and small scale limits}\label{constraints}
The governing equation for fluid motion, the Navier--Stokes--equation, imposes constraints that a complete model of turbulence needs to satisfy. 

One of those constraints is the symmetry $p(-u,-r)=p(u,r)$ \cite{Yakhot:1998:pdf}. It is fulfilled by Yakhot's original model but violated by both terms of the large--scale extension proposed in \cite{Me:2025:LargeScaleYakhot}. Restoring this symmetry hence requires adjustments to the parameter $D$ and the function $c(r)$. 

Out of these two parameters, only $D$ is discussed hereinafter\footnote{The function $c(r)$ was introduced to cover large--scale effects and was designed with the explicit aim of not having influence on small scales. It is hence not considered in this paper with its focus on small scales.}. For the $D$--term to satisfy the symmetry condition it has to be made a function of scale $r$ with $D(-r)=-D(r)$ \cite{Me:2025:LargeScaleYakhot}. It is convenient to separate the scale--dependence from the large--scale condition (\ref{modelParameterization}) by defining 
\begin{equation}
	D(r) \; = \; -\, 2\, A \, d(r) \label{dOfRdef}
\end{equation}
where, to fulfill symmetry and large--scale boundary conditions, $d(r)$ needs to satisfy:
\begin{eqnarray}
	d(-r) \; & = & - d(r) \nonumber \\
	d(r=1) \; & = & \; 1 \label{dSpec}
\end{eqnarray}

Another important constraint is Kolmogorov's famous four--fifth law (\ref{fourfifthWithDimensions}). Re--writing it  in the dimensionless variables $u$ and $r$ as defined in eq. (\ref{uAndR}) introduces the dimensionless rate of energy dissipation $\epsilon = \mathcal{E} \frac{L}{\sigma^3}$ and the (slightly modified\footnote{The Reynolds number is usually defined in terms of the mean velocity rather than the rms $\sigma$.}) Reynolds number $Re = \frac{\sigma L}{\nu}$:
\begin{eqnarray}
	S_3(r) \; & = & \; - \, \frac{4}{5} \, \epsilon \, r \; + \; \frac{6}{Re} \, \frac{\partial}{\partial r} \, S_2(r). \label{fourfifthDimensionless} 
\end{eqnarray}

In the limit $r \rightarrow 0$, the longitudinal structure function of order $n$ scales with $r^n$ in leading order as can be seen by expanding the velocity increment in a Taylor series:
\begin{eqnarray}
	S_n(r) \; =  \; \left< \, \left( w(x+r) - w(x) \right)^n \, \right> \; = \;  \left< \, \left( \frac{\partial w}{\partial x} \right)^n \, \right> \, r^n \; + \; \mathcal{O}(r^{n+1}). \label{viscousScaling}
\end{eqnarray}\
This can be used to derive an explicit expression for the second order structure function in the limit of small scales. Rearranging eq. (\ref{fourfifthDimensionless}) yields
\begin{eqnarray} 
	 \frac{\partial}{\partial r} \, S_2(r) \; & = & \;  \underbrace{ \, \frac{Re}{6} \, S_3(r) \, }_{\mathcal{O}(r^3)} \; + \; \underbrace{ \, \frac{2}{15} \, \epsilon \, Re \, r \,}_{\mathcal{O}(r)} 
	\;\; \underset{r \rightarrow 0}{=} \;\; \frac{2}{15} \, \epsilon \, Re \, r 
\end{eqnarray}
and accordingly:
\begin{eqnarray}
	\lim_{r \rightarrow 0} \, S_2(r) \; & = & \; \frac{1}{15} \, \epsilon \, Re \, r^2. \label{smallScaleS2Limit}
\end{eqnarray}

\section{Experimental small-scale term for even order}\label{evenSmallScaleTerm}

With a scale dependent $D$--term according to eq. (\ref{dOfRdef}), expression (\ref{largeScaleExtendedSun}) for the structure functions becomes:
\begin{eqnarray}
	\frac{\partial}{ \partial r} S_{n}(r) \; & = & \; \frac{\zeta_{n}}{r} \, S_{n}(r) \, + \, z_{n} \, c(r) \, S_{n-1}(r) \, - \, 2(n-1) \,\zeta_n \, d(r) \, S_{n-2}(r), \label{scaleDependentDtermSuns}
\end{eqnarray}
where $c(r)$ is given by eq. (\ref{cDefinition}).

In accordance with the approach taken for the large--scale model \cite{Me:2025:LargeScaleYakhot}, we seek for extensions of the governing equations in the form of elementary functions as possible. In this spirit, we assume the function $d(r)$ to be universal:

\begin{hyp}[H\ref{dHypothesis1}] \label{dHypothesis1}
	The function $d(r)$ is independent of order $n$.
\end{hyp}

Hypothesis H\ref{dHypothesis1} allows to address the question whether a scale--dependent function $d(r)$ in eq. (\ref{scaleDependentDtermSuns}) is sufficient to obtain a correct full--scale description of the $S_n(r)$. For this purpose, the function $d(r)$ is implied from experimental data for structure functions of different order. If eq. (\ref{scaleDependentDtermSuns}) was a complete full--scale model of turbulence under hypothesis H\ref{dHypothesis1}, functions $d(r)$ implied from structure functions of different order $n$ would collapse into a single curve. If, on the other hand, these implied $d(r)$ exhibit pronounced differences, it can be conclude that (\ref{scaleDependentDtermSuns}) comprises additional terms.

The function $d(r)$ is implied from experimental data according to
\begin{eqnarray}
	d_n(r) \; = \; \frac{\frac{\zeta_{n}}{r} \, S_{n}(r)  \, + \, z_{n} \, c(r) \, S_{n-1}(r) \, - \, \frac{\partial}{ \partial r} S_{n}(r)}{2(n-1) \,\zeta_n \, S_{n-2}(r)}, \label{impliedDofR}
\end{eqnarray}
where the subscript $n$ was introduced to distinct these implied from the actual model function $d(r)$.

\begin{figure}[ht]
	\centering
		\includegraphics[width=0.5\textwidth]{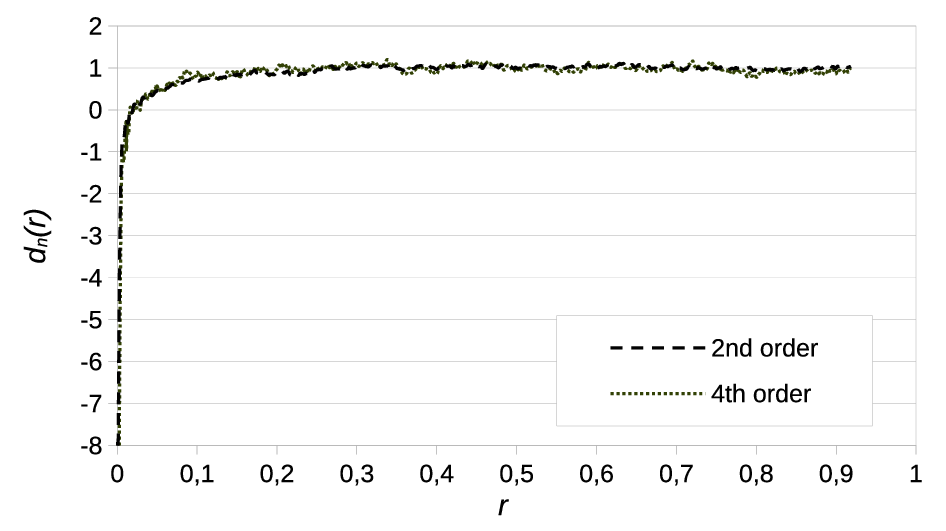}\hfill
		\includegraphics[width=0.5\textwidth]{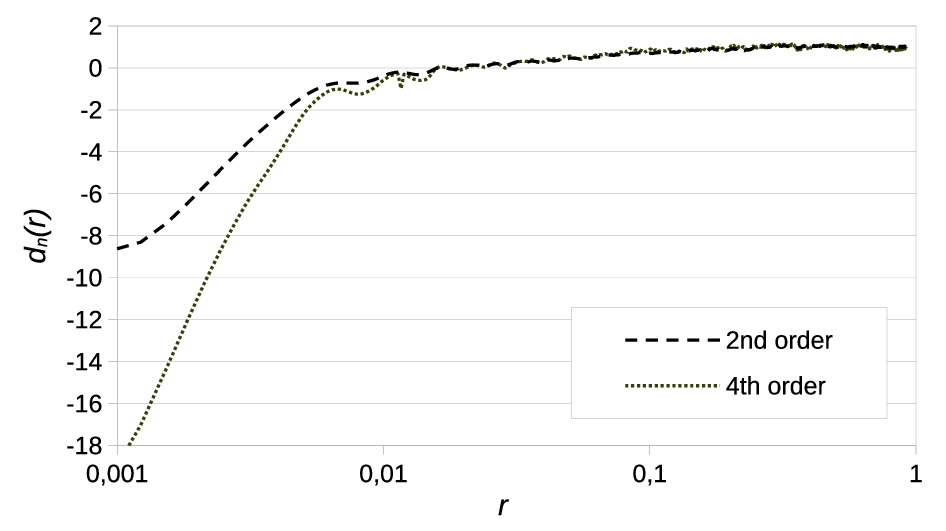}
	\caption{Functions $d_n(r)$ as defined in eq. (\ref{impliedDofR}) implied from experimental data of structure functions of order two (dashed lines) and four (dotted lines) in linear (left) and log--linear (right) scale.}
	\label{largeScaleImpliedDofR}
\end{figure}

Figure \ref{largeScaleImpliedDofR} shows the $d_n(r)$ for order two and four in linear (left) and log--linear (right) scale. The results differ significantly for small scales and it is concluded that an adaption of the $D$--term, albeit necessary to restore symmetry, is not sufficient to obtain a full--scale model of turbulence and that at least one additional term has to be introduced in (\ref{scaleDependentDtermSuns}).

In light of this result and the observation that the implied $d_n(r)$ are close to one over a wide range of scales, see figure \ref{largeScaleImpliedDofR}, the further analysis is based on an even stricter hypothesis on the function $d(r)$. Assuming a strict separation of large and small scale effects, we postulate that $d(r)$ does not contribute to the small scale dynamics at all:

\begin{hyp}[H\ref{dHypothesis2}] \label{dHypothesis2}
	The function $d(r)$ retains its large--scale value of one for most scales and approaches zero just fast enough as to not have any influence on the dynamics at small scales.
\end{hyp}

Under hypothesis H\ref{dHypothesis2}, the necessary additional term in (\ref{scaleDependentDtermSuns}) can approximately be implied from experimental data. We denote this term by $R_n(r)$ and define it as the residual of the difference of the left and right hand sides of eq. (\ref{scaleDependentDtermSuns}) where $d(r)=1$:
\begin{eqnarray}
	R_n(r) = \frac{\partial}{ \partial r} S_{n}(r) - \left\{ \frac{\zeta_{n}}{r} S_{n}(r) + z_{n} c(r) S_{n-1}(r) - 2(n-1) \zeta_n S_{n-2}(r)  \right\}. \label{resDefinition}
\end{eqnarray}

\begin{figure}[ht]
	\centering
		\includegraphics[width=0.5\textwidth]{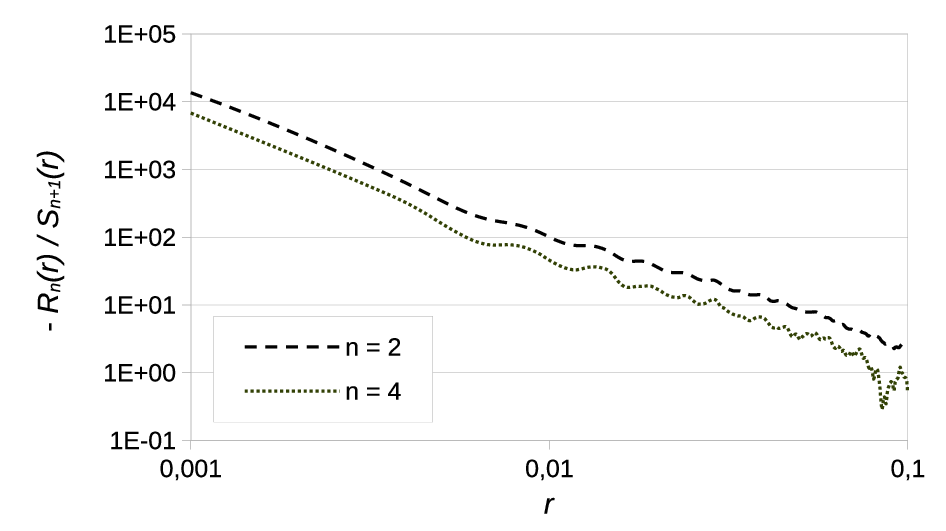}\hfill
		\includegraphics[width=0.5\textwidth]{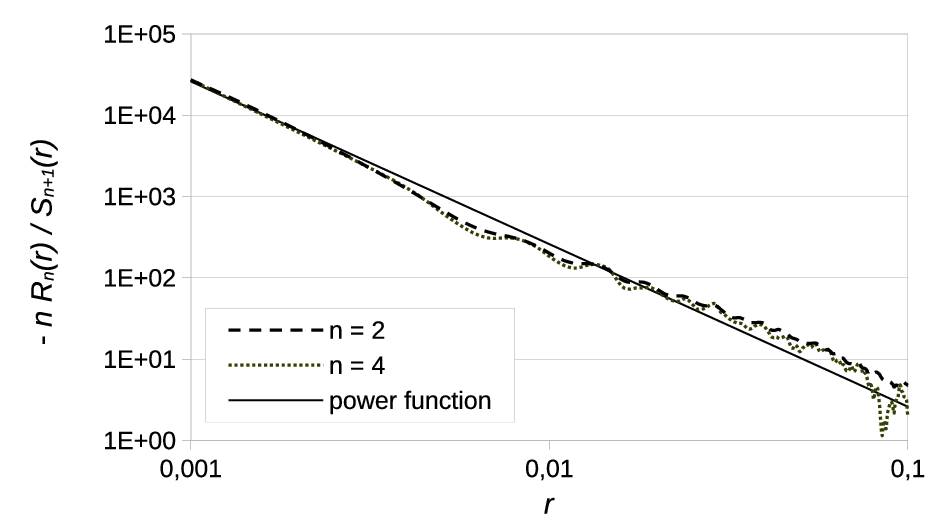}
	\caption{The residuals $R_n(r)$ as defined in eq. (\ref{resDefinition}) for orders $n=2$ (dashed lines) and $4$ (dotted lines) compensated by the structure functions $S_{n+1}$ (left) and additionally scaled by order $n$ (right). The straight line in the right graph shows a fit according to eq. (\ref{rnDef}). }
	\label{compensatedResiduals}
\end{figure}

In an attempt to retain the general structure of equation (\ref{scaleDependentDtermSuns}), we seek to express $R_n(r)$ as the product of an elementary function and a structure function of some order. Relating the residuals with structure functions of different orders, a particular result stands out: The ratios of $R_n(r)$ and the structure functions of order $(n+1)$ follow power laws in $r$, see figure \ref{compensatedResiduals}. What is more, when multiplied with order $n$ these compensated residuals collapse into a single function which is in good approximation described by a power law in $r$ with integer exponent $-2$:
\begin{eqnarray}
	- \, n \, \frac{R_n(r)}{ S_{n+1}(r)} \; = \; \frac{\tau}{r^2}. \label{rnDef}
\end{eqnarray}
For the data set at hand, we obtain a value of $\tau \approx  0.026$ from a fit to the compensated residuals. 

The fact that the power law (\ref{rnDef}) is in good approximation observed also for the smallest scales resolved by the experiment supports the initial assumption that in this range $d(r)$ is close to its large--scale value of one. Deviations from that level and the eventual decline towards zero can thus be expected to set in only at dissipative length scales which are not fully resolved by the experiment.

With expression (\ref{rnDef}) for the $R_n(r)$, the full equation for even structure functions (of orders two and four) is obtained as:
\begin{eqnarray}
	\frac{\partial}{ \partial r} S_{n}(r) \; = \; & - &  \frac{1}{n}\, \frac{\tau}{r^2} \, S_{n+1}(r) \, +  \, \frac{\zeta_{n}}{r} \, S_{n}(r) \nonumber \\
		& + & z_{n} \, c(r) \, S_{n-1}(r) \, - \, 2(n-1) \, \zeta_n \, d(r) \, S_{n-2}(r). \label{fullSnEqForEvenN}
\end{eqnarray}
It is conceivable that relations (\ref{rnDef}) and (\ref{fullSnEqForEvenN}) can be generalized to even $n>4$, but the verification of this assumption requires data with higher statistical resolution and has to be left for future research. 

Yet, it should not be assumed that equation (\ref{rnDef}) is an exact relation. Rather, preliminary results (see appendix \ref{prelimResults}) indicate that it becomes less accurate for larger scales and lower Reynolds numbers. These inaccuracies have only insignificant impact for the description of $\partial_r S_{n}(r)$ via (\ref{fullSnEqForEvenN}) as they are dampened by the prefactor $\propto r^{-2}$ , but would be amplified by a factor $\propto r^{2}$ if the equation was solved for $S_{n+1}(r)$. Relation (\ref{fullSnEqForEvenN}) can thus serve as a building  block for a model of even order structure functions, but can most likely not be used to model odd order moments.

\begin{figure}[ht]
	\centering
		\includegraphics[width=0.5\textwidth]{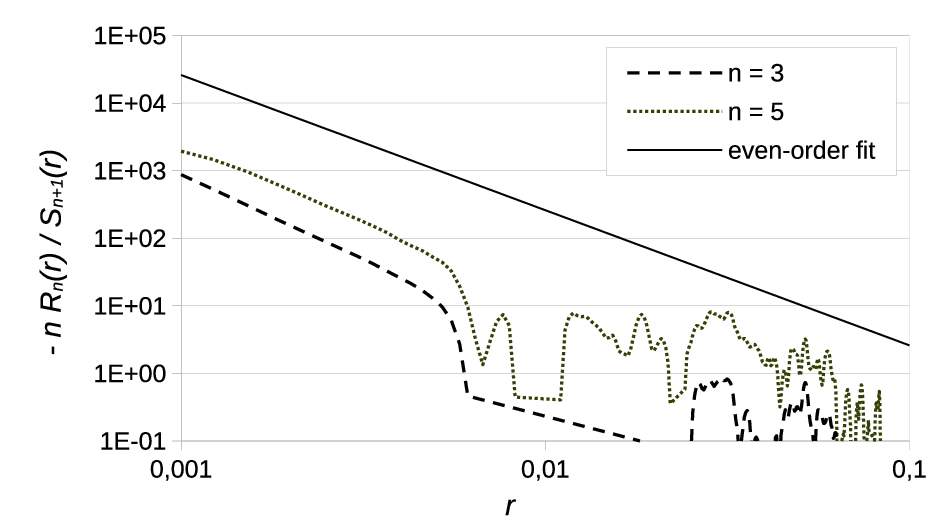}\hfill
		\includegraphics[width=0.5\textwidth]{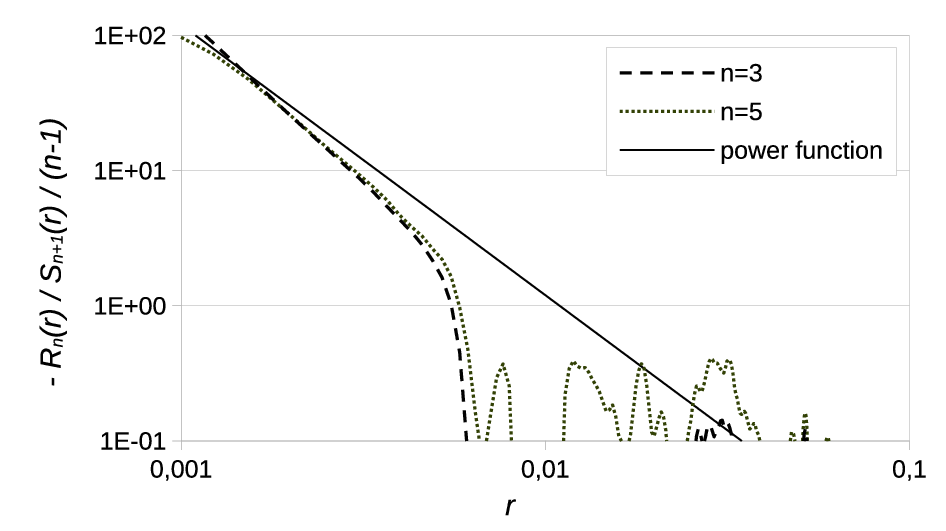}
	\caption{Compensated and scaled residuals $R_n(r)$ for orders $n=3$ (dashed lines) and $n=5$ (dotted lines) with power law fits (solid lines). Left graph: Residuals scaled with order $n$ in comparison to the power law fit of even--order residuals according to eq. (\ref{rnDef}). Right graph: Results for alternative scaling with $1/(n-1)$ in comparison to a power--function fit of second order (the fitted pre--factor does not coincide with the value of $\tau$ found for even order).}
	\label{compensatedOddResiduals}
\end{figure}

A complete full--scale model of turbulence hence requires a further relation for odd order. Unfortunately, odd order residuals $R_n(r)$ are found to not follow relation (\ref{rnDef}). Figure \ref{compensatedOddResiduals} shows the residuals for orders $n=3$ and $n=5$ compensated by the structure functions of or orders $4$ and $6$. When multiplied with order $n$, analogous to the approach for even order, the residuals do not collapse into a universal function. What is more, the power--law in $r$ found for even order, eq. (\ref{rnDef}) with $\tau \approx  0.026$, fails to even only describe the order of magnitude of the scaled and compensated odd order residuals. 

Interestingly enough, the compensated odd order residuals still show a similar dependence on scale $r$ and in fact can be brought into agreement. However, while even order residuals need to be multiplied by  $n$, the compensated odd order moments have to be divided by $(n-1)$ in order to exhibit universal behaviour. Also, the resultant curve cannot be described by a power law, but exhibits a more complex dependence\footnote{
When drawing conclusions from the results presented in figure \ref{compensatedOddResiduals} it needs to be kept in mind that these are derivatives of odd order structure functions. Owing to the fact that fluctuations with opposite sign partially cancel out, these are to a larger extent affected by noise than even order moments and their derivatives.
}
on scale $r$ (figure \ref{compensatedOddResiduals}, right).

With the results presented above, a well defined and complete full--scale version of the model equations can only be given for even order structure functions. To make matters worse, these equations are not closed as the small-scale term in (\ref{fullSnEqForEvenN}) comprises the odd order structure function $S_{n+1}$. 

For orders two and three, however, the four--fifth law (\ref{fourfifthDimensionless}) provides an additional relation that allows to close the equations and to derive closed--form expressions for the structure functions $S_2$ and $S_3$. These will in the following be derived for $S_2$ in the small--scale limit first before in a second step the description will be extended to cover also large scales.

\section{The second order structure function in the limit of small scales}\label{secondOrderSmallScales}

The derivation of the full scale model (\ref{fullSnEqForEvenN}) crucially depends on hypothesis H\ref{dHypothesis2}. In order to verify (or falsify) this premise, in particular the assumption that $d(r)$ does not contribute to the small--scale dynamics, the model is next solved under the assumption of a vanishing $d$--term and compared with experimental data for $S_2$. In case the model does indeed describe the structure function for small scales, this analysis would not only support the initial assumption of $d(r)$ approaching zero for $r \rightarrow 0$, but should  also allow to narrow down the typical length scale at which this transition takes place.

For vanishing $d$, equation (\ref{fullSnEqForEvenN}) for the second order structure function reduces to:
\begin{eqnarray}
	\frac{\partial}{ \partial r} S_{2}(r) \;  = \;  -  \frac{1}{2} \, \frac{\tau}{r^2} \, S_{3}(r) \, +  \, \frac{\zeta_{2}}{r} \, S_{2}(r). \label{fullSmallScaleSTwo}
\end{eqnarray}

Using the four--fifth law (\ref{fourfifthDimensionless}) to replace $S_{3}(r)$ we obtain a closed equation for the second order structure function:
\begin{eqnarray}
	\frac{\partial}{ \partial r} S_{2}(r) \; & = & \;   \frac{2 \, Re}{15} \, \epsilon \, \rho^2 \, \frac{g(r)}{r} \, +  \, \zeta_{2}\, \frac{g(r)}{r} \, S_{2}(r)  \label{closedSmallScaleStwo}
\end{eqnarray}
where
\begin{eqnarray}
	g(r) \; & = & \; \frac{r^2}{r^2 \, + \, \rho^2}, \nonumber \\
	\rho^2 \; & = & \; \frac{3 \, \tau}{Re}. \label{gAndRhoDef} 
\end{eqnarray}
As apparent from the definition of the function $g(r)$, the parameter $\rho$ constitutes a characteristic length scale of the model. For the data set at hand the Reynolds number is approximately $Re \approx 2.3 \cdot 10^5$. With the experimental value $\tau \approx 0.026$ this yields a value of $\rho \approx 5.8 \cdot 10^{-4}$.

The general solution of (\ref{closedSmallScaleStwo}) is:
\begin{eqnarray}
	S_2(r) \; = \; K_2 \, \left( \, r^2 \, + \, \rho^2 \, \right)^\frac{\zeta_2}{2} \, - \, \frac{2 \, Re }{15 \, \zeta_2} \, \epsilon \, \rho^2.
\end{eqnarray}

The trivial boundary condition $S_2(0) = 0$ is not readily fulfilled and can be used to determine the integration constant $K_2$. The full solution for $S_2(r)$ for $d=0$ is:
\begin{eqnarray}
	S_2(r) \; & = & \; \frac{2 \, Re }{15 \, \zeta_2} \, \epsilon \, \rho^2 \, \left\{ \, \left( \, 1  \, + \, \frac{r^2}{\rho^2} \, \right)^\frac{\zeta_2}{2} \, - \, 1 \, \right\}. \label{smallScaleS2Formula}
\end{eqnarray}

\begin{figure}[ht]
	\centering
		\includegraphics[width=0.75\textwidth]{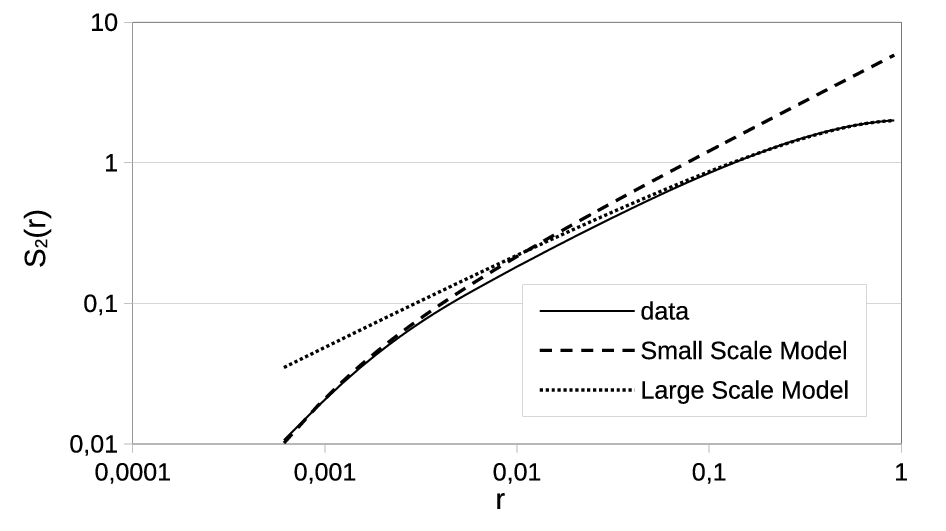}
	\caption{Second order structure function determined from experimental data (straight line) in comparison to the small (dashed line) and the large scale models (dotted line) as given by equations (\ref{smallScaleS2Formula}) and (\ref{ExtS2Solution}), respectively.}
	\label{smallScaleS2ModelFit}
\end{figure}

Figure \ref{smallScaleS2ModelFit} shows a comparison of the small--scale model (\ref{smallScaleS2Formula}) with experimental data. The model fits the smallest scales up to the transition to the inertial range well. As expected, it then exhibits increasingly larger deviations from the data in the inertial range.

It is easily verified that (\ref{smallScaleS2Formula}) has the expected asymptotics for $r \rightarrow 0$. For $r \ll \rho$ we can approximate
\begin{eqnarray}
	\left( \, 1  \, + \, \frac{r^2}{\rho^2} \, \right)^{\frac{\zeta_2}{2}} \; \approx \; 1 + \frac{\zeta_2}{2} \,  \frac{r^2}{\rho^2} \label{muchSmallerThanRhoProxy}
\end{eqnarray}
and in accordance with eq. (\ref{smallScaleS2Limit}) obtain:
\begin{eqnarray}
	\lim_{r \rightarrow 0} \, S_2(r) \; = \; \frac{Re}{15} \, \epsilon \, r^2 .   \label{fourFifthConfirmed}
\end{eqnarray}

This result is important not only in that it proves consistency of the model with the small--scale scaling law (\ref{smallScaleS2Limit}), thus providing strong support for hypothesis H\ref{dHypothesis2}, but also in that it identifies $\rho$ as the scale at which the convergence against that limit sets in. We can hence identify $\rho$ as the scale at which $d(r)$ decreases to zero.

The limit $r \gg \rho$ yields another important result for the characteristic scale $\rho$. In this limit, terms of order unity in the brackets can be neglected and the solution (\ref{smallScaleS2Formula}) for $S_2(r)$ reduces to the classical scaling--law in $r$ with exponent $\zeta_2$:
\begin{eqnarray}
	S_2(r) \;	& \approx & \;  \frac{2 \, Re }{15 \, \zeta_2} \, \epsilon \, \rho^2 \, \left( \, \frac{r^2}{\rho^2} \, \right)^\frac{\zeta_2}{2} 
						\; = \;  \frac{2 \, Re }{15 \, \zeta_2} \, \epsilon \, \rho^{2-\zeta_2} \, r^{\zeta_2} .  \label{smallScaleGoesScaling} 
\end{eqnarray}

Comparing (\ref{smallScaleGoesScaling})  with the large--scale model (\ref{ExtS2Solution}), we can identify and equal the pre--factors of the scaling terms:
\begin{eqnarray}
	\frac{2 \, Re }{15 \, \zeta_2} \, \epsilon \, \rho^{2-\zeta_2}  \; \approx \;  \frac{2}{1 \, - \, \zeta_2}.  \label{firstOrderScalingFactors}
\end{eqnarray}
This yields a first order expression for the length scale $\rho$:
\begin{eqnarray}
	\rho^{2-\zeta_2} \; & \approx & \;   \frac{15 \, \zeta_2}{1 \, - \, \zeta_2} \, \frac{1}{ \epsilon \, Re }.  \label{firstOrderRho}
\end{eqnarray}

The importance of this result lies in the fact that it fixes the newly introduced parameter $\rho$ (or, equivalently, $\tau$) and allows to express it in terms of the Reynolds number and the parameters of the large--scale model. For the data set at hand with $Re \approx 2.3 \cdot 10^5$ this yields a value of $\rho \approx 6.1 \cdot 10^{-4}$, in reasonable agreement with the experimental value of $5.8 \cdot 10^{-4}$.

The limits for small and large scales, eqs. (\ref{fourFifthConfirmed}) and (\ref{smallScaleGoesScaling}), also shed light on the significance of the length scale $\rho$. These two limiting cases are power laws in $r$ with exponents of $2$ and $\zeta_2 \approx 0.7$, i.e. convex and concave functions, respectively. The point $r_0$ at which $S_2(r)$ passes over from one to the other is characterized by a vanishing second derivative, $\partial^2_r S_2(r_0) = 0$, and is obtained from (\ref{smallScaleS2Formula}) as:
\begin{eqnarray}
	r_0^2 \; = \; \frac{1}{1-\zeta_2} \, \rho^2.
\end{eqnarray}
Up to a factor of order unity, the scale $\rho$ hence marks the transition from the viscous $r^2$ to the inertial range $r^{\zeta_2}$ scaling.

The relation between $\rho$ and the Kolmogorov microscale can further be elucidated by using the Kolmogorov 1941 model approximation $\zeta_2 \approx 2 / 3$. With it, relation (\ref{firstOrderRho}) for $\rho$ can further be simplified to:
\begin{equation}	
	\rho \; \approx \; \left( \, \frac{\epsilon}{30} \, \right)^{-\frac{3}{4}} \, Re^{-\frac{3}{4}}
\end{equation}
For perspective, expressing the Kolmogorov microscale in dimensionless units yields:
\begin{equation}	
	\frac{\eta}{L} \; = \; \epsilon^{-\frac{1}{4}} \, Re^{-\frac{3}{4}}
\end{equation}
To first order, $\rho$ and $\eta$ exhibit the same dependence on the Reynolds number, albeit with prefactors that differ by roughly one order of magnitude. This ties in with the observation that $\rho$ marks the crossover point from inertial to dissipative scaling regimes, whereas the much smaller Kolmogorov microscale defines the length scale at which dissipation finally dominates.

\section{A full scale model for second and third order}\label{fullScaleModel}

To complete the specification of the full--scale equation (\ref{fullSnEqForEvenN}), the function $d(r)$ still needs to be specified. From the analyses presented above it is known that it
\begin{itemize}
	\item is an odd function, i.e. $d(-r)=-d(r)$, 
	\item decreases to zero as $r$ approaches zero and
	\item quickly converges against $1$ for $r > \rho$.
\end{itemize}
These are the general characteristics of a sigmoid function with characteristic length scale $\rho$. Being a class of functions, the characterization as a sigmoid does not uniquely fix $d(r)$ so that an additional assumption needs to be taken. 

An aspect worth considering in that context is analytical tractability of the model. It is best if $d(r)$ is chosen to be\footnote{
	Strictly speaking, this choice is wrong in that $d(r=1) \neq 1$. The correct large--scale level of $1$ could be enforced via a normalization factor, but since $\rho$ is small, this factor is close to one and omitted for the sake of simplicity.
	}\textsuperscript{,}\footnote{
		The specific choice of the sigmoid has an only minor impact. In Appendix \ref{fullPrescisionS2} we report on results of a test calculation where the hyperbolic tangent is used to model $d(r)$.
}
\begin{eqnarray}
	d(r) \; =  \; \frac{r}{\sqrt{r^2 + \rho^2}} \label{dDef}
\end{eqnarray}
which implies that $d^2(r) = g(r)$. 

\subsection{Equations for second order}
The equation for the second order moment
\begin{eqnarray}
	\frac{\partial}{ \partial r} S_{2}(r) \; = \;  - \, \frac{1}{2}\, \frac{\tau}{r^2} \, S_{3}(r) \, +  \, \frac{\zeta_{2}}{r} \, S_{2}(r) \, - \, 2 \, \zeta_2 \, d(r), \label{fullS2Eq}
\end{eqnarray}
can be closed with the help of the four--fifth law (\ref{fourfifthDimensionless}). In analogy to the small--scale model (\ref{closedSmallScaleStwo}) we obtain:
\begin{eqnarray}
	\frac{\partial}{ \partial r} S_{2}(r) \;	=  \;   \frac{2 \, Re}{15} \, \epsilon \, \rho^2 \, \frac{g(r)}{r} \, +  \, \zeta_{2}\, \frac{g(r)}{r} \, S_{2}(r) \, - \, 2 \, \zeta_2 \, g(r)\, d(r) \label{closedFullScaleStwo}
\end{eqnarray}
with $g(r)$ and $\rho$ as in (\ref{gAndRhoDef}). 

It needs to be mentioned that closing eq. (\ref{fullS2Eq}) with the help of the four--fifth law introduces an error into the model: The derivative of $S_2$ at the system length scale is not zero any more. This violation is caused by the first term on the right--hand side of eq. (\ref{closedFullScaleStwo}) which does not vanish at $r=1$ (whereas the second and third term cancel out at $r=1$). 

The term originates from the linear term in the four--fifth law (\ref{fourfifthDimensionless}) which is a valid approximation in the inertial range only. A model consistent with the large--scale condition can be obtained by combining the four--fifth law with the large--scale model (\ref{ExtS3Solution}). On the other hand, as outlined in appendix \ref{fullPrescisionS2}, this correction has an impact of only second order. For the clarity of presentation we will thus proceed without this correction.

With $g(r)$ and $d(r)$ as defined in (\ref{gAndRhoDef}) and (\ref{dDef}), the solution of eq. (\ref{closedFullScaleStwo}) is:
\begin{eqnarray}
	S_2(r) \; = \;  K_2 \left( r^2 + \rho^2 \right)^\frac{\zeta_2}{2} \; - \; \frac{2 \, Re}{15 \, \zeta_2} \, \epsilon \, \rho^2 \; - \; 2 \frac{\zeta_2}{1-\zeta_2} \frac{r^2}{\sqrt{r^2+\rho^2}} \nonumber \\ \; - \; 4 \frac{\zeta_2}{1-\zeta_2^2}  \frac{\rho^2}{\sqrt{r^2+\rho^2}} \, .
\end{eqnarray}

As for the small--scale approximation, the trivial boundary condition $S_2(0)=0$ can be used to determine the integration constant $K_2$. With this we finally obtain:
\begin{eqnarray}
	S_2(r) \;  =  \;  \frac{2 \, Re}{15 \, \zeta_2} \, \epsilon \, \rho^2 \, \left\{ \, \left( 1 + \frac{r^2}{\rho^2} \right)^\frac{\zeta_2}{2} - 1 \, \right\} \;
								\; - \;  2 \, \frac{\zeta_2}{1-\zeta_2} \, \frac{r^2}{\sqrt{r^2+\rho^2}} 	\nonumber \\
								\; + \; 4 \, \rho \, \frac{\zeta_2}{1-\zeta_2^2} \, \left\{ \, \left( 1 + \frac{r^2}{\rho^2} \right)^\frac{\zeta_2}{2} - \, \frac{\rho}{\sqrt{r^2+\rho^2}} \, \right\} \, . \label{fullScaleS2}
\end{eqnarray}

The length scale $\rho$ can be determined from the large--scale boundary condition $S_2(1)=2$. Numerically a value of $\rho = 5.74 \cdot 10^{-4}$ is found, in good agreement with the experimental value of $5.8 \cdot 10^{-4}$.

A more insightful expression for $\rho$ can be derived from the limit $r \gg \rho$ of (\ref{fullScaleS2}). By approximating $r^2+\rho^2 \approx r^2$, $\rho / r \approx 0$ and neglecting terms of order unity in the brackets, the large--scale limit of (\ref{fullScaleS2}) becomes:
\begin{eqnarray}
	S_2(r) \;  \approx \; \left\{ \, \frac{2 \, Re}{15 \, \zeta_2} \, \epsilon \, \rho^2 \,  + \, 4 \rho \frac{\zeta_2}{1-\zeta_2^2} \, \right\}  \, \left( \frac{r}{\rho}\right)^{\zeta_2} \; - \; \frac{2}{1-\zeta_2} \, \zeta_2 \, r. \label{S2LargeScaleProxyFromFullModel}
\end{eqnarray}

The approximation coincides with the large--scale model (\ref{ExtS2Solution}) with respect to the functional dependencies on scale $r$ and the also prefactor of the linear term. We can hence identify the prefactor of the scaling--term in (\ref{S2LargeScaleProxyFromFullModel}) with the corresponding factor in (\ref{ExtS2Solution}):
\begin{eqnarray}
	\left\{ \, \frac{2 \, Re}{15 \, \zeta_2} \, \epsilon \, \rho^2 \,  + \, 4 \, \rho \, \frac{\zeta_2}{1-\zeta_2^2} \, \right\} \, \rho^{-\zeta_2}  \;  = \;  \frac{2}{1 - \zeta_2}. \label{fullRho}
\end{eqnarray}

This expression is a straightforward extension of the first--order approximation (\ref{firstOrderScalingFactors}). For the data set at hand we numerically obtain $\rho = 5.72 \cdot 10^{-4}$, in agreement with the value of $5.74 \cdot 10^{-4}$ obtained from eq. (\ref{fullScaleS2}).

\begin{figure}[ht]
	\centering
		\includegraphics[width=0.5\textwidth]{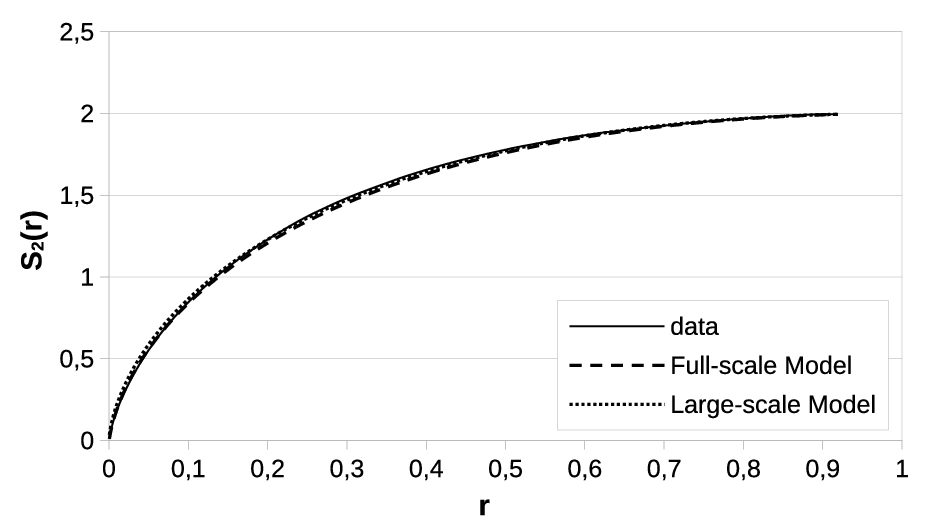}\hfill
		\includegraphics[width=0.5\textwidth]{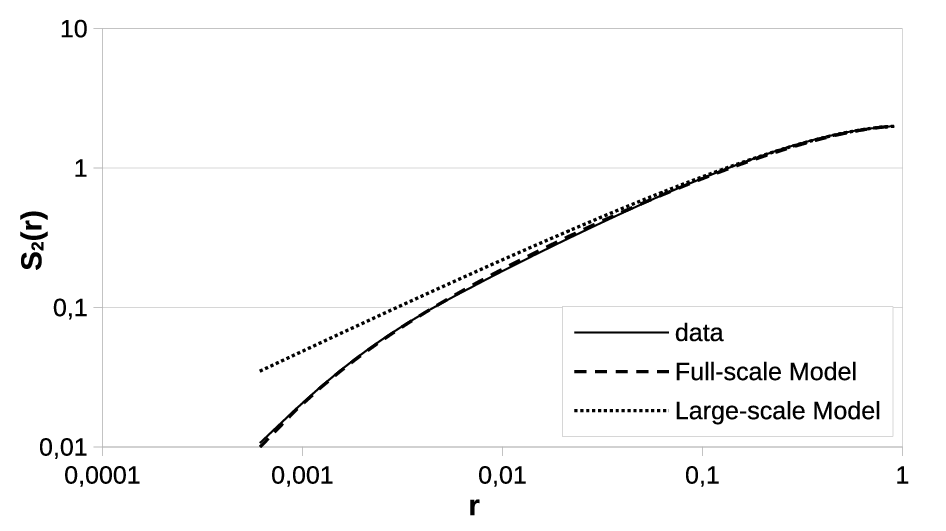}
	\caption{The second order structure function (solid line) in comparison with large--scale (dotted line) and full--scale (dashed line) models as given by equations (\ref{ExtS2Solution}) and (\ref{fullScaleS2}), respectively, in linear (left) and logarithmic (right) scale. The scale parameter $\rho$ of the model (\ref{fullScaleS2}) was set to a value of $\rho = 5.7 \cdot 10^{-4}$ according to eq. (\ref{fullRho}).}
	\label{fullScaleS2Fit}
\end{figure}

Figure \ref{fullScaleS2Fit} shows a comparison of the full--scale model (\ref{fullScaleS2}) with experimental data and the large--scale model. In both linear as well as logarithmic scale, the model exhibits good agreement with experimental data over the entire range of length scales resolved by the experiment.

\subsection{Equations for third order}
The model for $S_2(r)$ can also be used to derive an explicit model for the third order structure function. This can be achieved with the help of the four--fifth law (\ref{fourfifthDimensionless}), which provides an exact expression for the third order structure function in the dissipation and inertial range. However, it has the obvious drawback that the large--scale boundary condition $S_3(r=1)=0$ is not fulfilled. A model consistent with this condition can be obtained by combining the four--fifth law (\ref{fourfifthDimensionless}) with the large scale model (\ref{ExtS3Solution}):
\begin{eqnarray}
	S_3(r) \; & = & \; - \, \frac{4}{5} \, \epsilon \, r \, \left\{ \, 1 \, - \frac{F(r)}{F(1)} \, \right\} \; + \; \frac{6}{Re} \, \frac{\partial}{\partial r} \, S_2(r). \label{largeScaleFourFifth}
\end{eqnarray}

The $S_2$--term is significant for small scales only and it is therefore sufficient to consider the small-scale expression (\ref{smallScaleS2Formula}) for $S_2(r)$ when calculating it. Inserting the result for $\partial_r S_2$ into (\ref{largeScaleFourFifth}) yields:
\begin{eqnarray}
		S_3(r) \; & \approx & \; - \, \frac{4}{5} \, \epsilon \, r \, \left\{ \, 1 \, - \frac{F(r)}{F(1)} \, - \, \left( 1 \, + \, \frac{r^2}{\rho^2} \right)^{-1 + \frac{\zeta_2}{2} } \, \right\}. \label{fullScaleS3}
\end{eqnarray}

\begin{figure}[ht]
	\centering
		\includegraphics[width=0.5\textwidth]{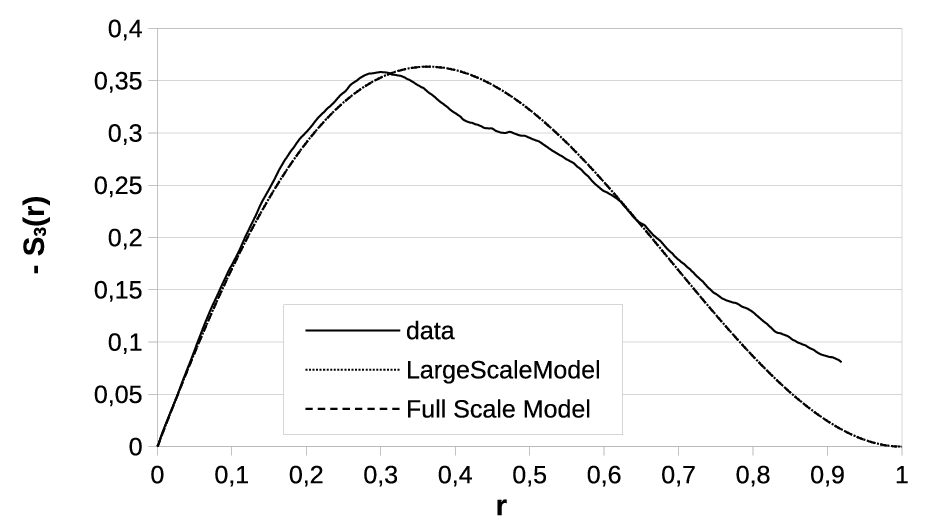}\hfill
		\includegraphics[width=0.5\textwidth]{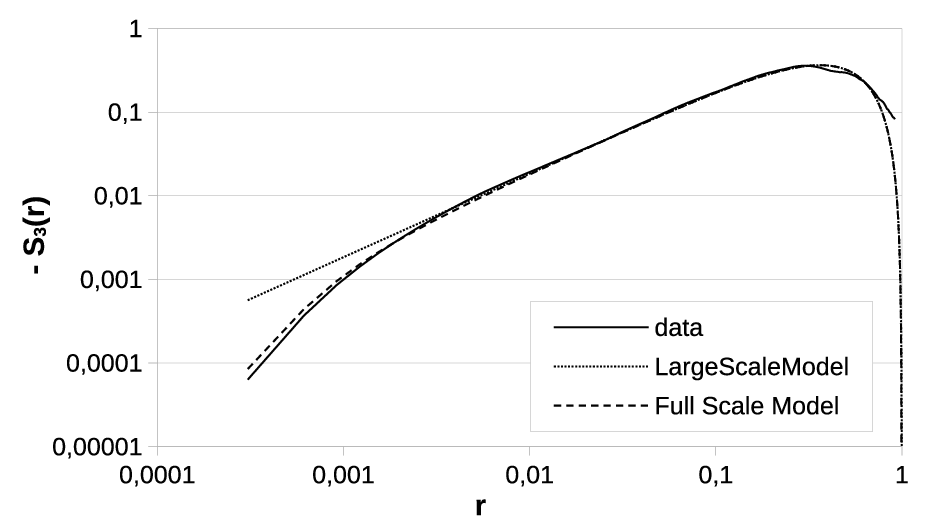}
	\caption{Third order structure function determined from experimental data (straight line) in comparison to full (dashed line) and large scale models (dotted line) as given by equations (\ref{fullScaleS3}) and (\ref{ExtS3Solution}), respectively in linear (left) and logarithmic (right) scale.}
	\label{fullScaleS3ModelFit}
\end{figure}

A comparison of the model (\ref{fullScaleS3}) with experimental data for $S_3(r)$. As can be seen in figure \ref{fullScaleS3ModelFit}, the model replicates the data well, from the smallest dissipative scales through to the upper end of the inertial range. 

We conclude this section with a discussion of the small--scale asymptotics of the model. For scales $r \ll \rho$ the large--scale $F(r)$--term, being of order $r{1+\zeta_2}$, can be neglected and the power in $r$ can be approximated as
\begin{eqnarray}	
	\left( \, 1  \, + \, \frac{r^2}{\rho^2} \, \right)^{-1 + \frac{\zeta_2}{2}} \; \approx \; 1 - \left( 1 - \frac{\zeta_2}{2} \right) \,  \frac{r^2}{\rho^2}, \nonumber
\end{eqnarray}
and we obtain from eq. (\ref{fullScaleS3}):
\begin{eqnarray}	
	\lim_{r \rightarrow 0} \, S_3(r) \; & = & \; - \, \frac{4}{5} \, \epsilon \, \left( 1 - \frac{\zeta_2}{2} \right) \, \frac{r^3}{\rho^2}.
\end{eqnarray}
The model is thus consistent with the expected small--scale scaling $\propto r^3$, see eq. (\ref{viscousScaling}).

\section{Summary and Discussion}\label{discussion}

The main result presented in this paper is the experimental relation (\ref{rnDef}) between the spatial derivatives of even order structure functions and the next higher odd moment. This relation completes the differential equation of Yakhot's model for structure functions of even order across all scales (experimentally verified for orders two and four), from the dissipation range through to the system scale:
\begin{eqnarray}
	\frac{\partial}{ \partial r} S_{n}(r) \; = \; & - &  \frac{1}{n}\, \frac{\tau}{r^2} \, S_{n+1}(r) \, +  \, \frac{\zeta_{n}}{r} \, S_{n}(r) \nonumber \\
		& + & z_{n} \, c(r) \, S_{n-1}(r) \, - \, 2(n-1) \, \zeta_n \, d(r) \, S_{n-2}(r). \label{evenOrderEquationAgain}
\end{eqnarray}

This result was derived under two assumptions on the function $d(r)$, hypotheses H\ref{dHypothesis1} and H\ref{dHypothesis2}. While little more than (the author's) physical intuition can serve as evidence for H\ref{dHypothesis1}, the arguably stronger hypothesis H\ref{dHypothesis2} is backed by the results of the model: The assumption of a vanishing $d$--term in the limit of small scales yields a model that replicates the expected small scale scaling law for the second order structure function and is in good agreement with experimental data in the dissipation range and the transition to the inertial range. What is more, the full model equations for $S_2$ and $S_3$ derived under this hypothesis show good agreement with experimental data over the entire range of scales, up to the system scale.

Still, in its present form the proposed model clearly is limited. Owing to the term of order $n+1$, eq. (\ref{evenOrderEquationAgain}) for even order $n$ is not self--contained. As furthermore an analogous expression for odd order has not yet been found, the model framework is incomplete and can only be closed for orders two and three (with the help of the four--fifth law), but not for arbitrary $n$.

Another dissatisfying aspect is the asymmetry between the results for even and odd order as apparent in figures \ref{compensatedResiduals} and \ref{compensatedOddResiduals}. This asymmetry makes it difficult to imagine how an extension of the more fundamental partial differential equation (\ref{largeScaleExtenssion}) for the probability density function $p(u,r)$ must be designed. It is also conceivable that this is not possible, i.e. that a complete model of structure functions cannot be formulated for the longitudinal component in isolation but only in a more general framework accounting for the interdependence of longitudinal and transversal increments as, for example, discussed in \cite{Boschung:2017}.

Yet, the analysis presented has a solid foundation. The scaling law (\ref{rnDef}) extends over a wide range of scales, from the dissipation well into the inertial range. It is likely an approximation only that becomes invalid for small Reynolds numbers, see appendix \ref{prelimResults}, but is still as clearly pronounced as hardly ever observed in experimental data, let alone with an integer scaling exponent.

Another notable result is the fact that the model does not comprise free parameters. The empirical scaling law (\ref{rnDef}) introduces one additional parameter into the model, $\tau$ or, equivalently, the length scale $\rho$. This scale marks the transition from the dissipative to the inertial scaling regime for the second order structure function and can be expressed as a function of the Reynolds number and the mean rate of energy dissipation.

The proposed extension of Yakhot's model hence yields a full--scale model of low order structure functions that is in good agreement with experimental data, replicates expected small--scale scaling laws and is fully determined by the large--scale characteristics of the flow (system length scale and Reynolds number) and the dimensionless mean rate of energy dissipation. We believe that this is a substantial result.

\subsection*{Acknowledgments} 
The author gratefully acknowledges fruitful discussions with J. Peinke and the generous provision of the high quality data sets by courtesy of B. Castaing and B. Chabaud.

\appendix

\section{Preliminary results for lower Reynolds numbers} \label{prelimResults}

The analyses presented in the main body of this paper (for the data set with $Re \approx 2.5\cdot 10^5$) have partially also been conducted for two data sets with lower Reynolds numbers. These have been measured in the same experimental facility, the cryogenic axisymmetric helium gas jet discussed in \cite{Chanal:2000:HeliumJet}, and have Reynolds numbers of approximately $2\cdot 10^4$ and $7\cdot 10^3$, respectively. 

\begin{figure}[ht]
	\centering
		\includegraphics[width=0.5\textwidth]{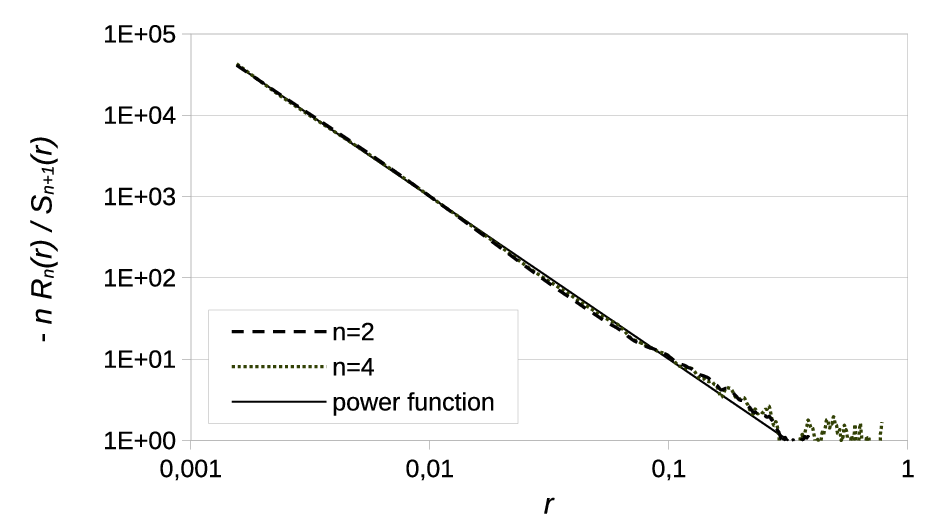}\hfill
		\includegraphics[width=0.5\textwidth]{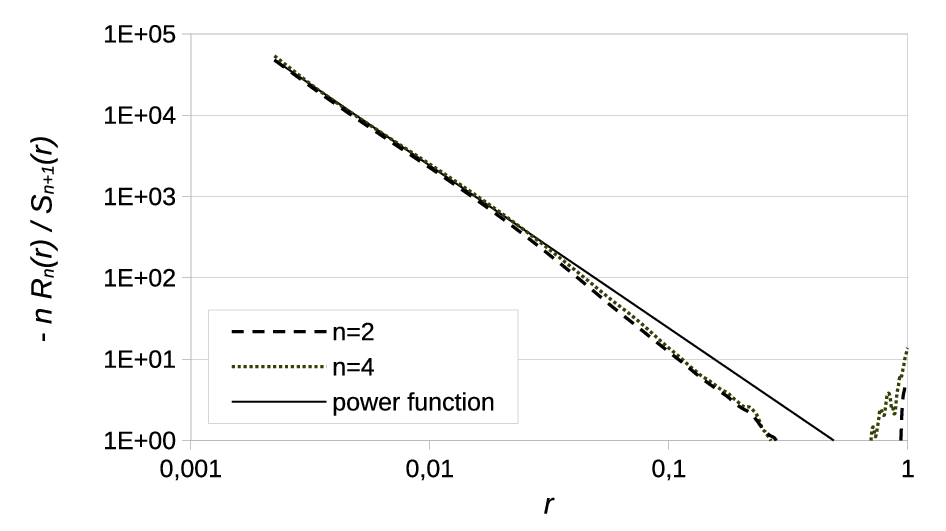}
	\caption{Even order compensated and scaled residuals $R_n(r)$ for measurements at Reynolds numbers of $2\cdot 10^4$ (left) and $7\cdot 10^3$ (right).}
	\label{residualsForOtherRe}
\end{figure}

The main purpose of these preliminary studies was to narrow down the range of validity of the proposed model. In a first step, it was examined whether the even order residuals $R_n(r)$ as defined by equation (\ref{resDefinition}) can be brought into agreement by scaling with order $n$ and compensation with the next higher order structure function $S_{n+1}(r)$. In analogy to the results presented in figure \ref{compensatedResiduals} we find that these scaled and compensated residual indeed collapse into universal functions for both measurements, see figure \ref{residualsForOtherRe}. Interestingly though, this universal function is described in good approximation by a power law with exponent $-2$ {\it only} for the data set with the higher Reynolds number, whereas for the data set with $Re \approx 7\cdot 10^3$ significant deviations from a simple power law are observed for larger scales.

\begin{figure}[H]
	\centering
		\includegraphics[width=0.5\textwidth]{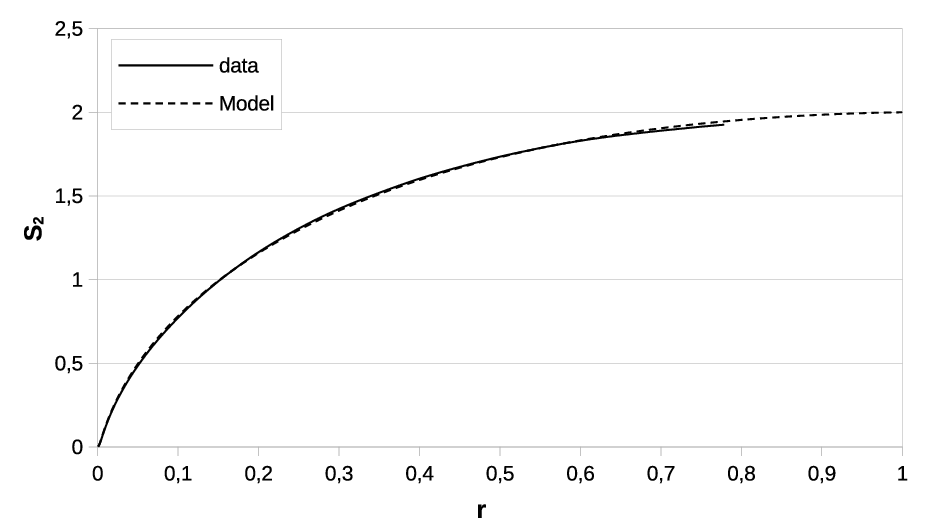}\hfill
		\includegraphics[width=0.5\textwidth]{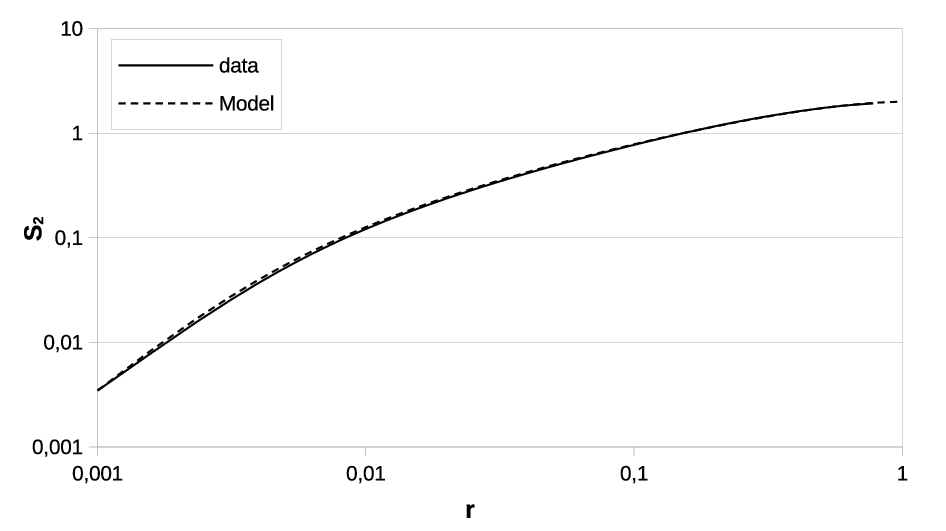}
	\caption{Second order structure function (straight lines) for the measurement with $Re=2\cdot 10^4$ in linear (left) and logarithmic (right) scale in comparison to the model prediction (dashed line) according to eq. (\ref{fullScaleS2}).}
	\label{s2forMediumRe}
\end{figure}

\begin{figure}[H]
	\centering
		\includegraphics[width=0.5\textwidth]{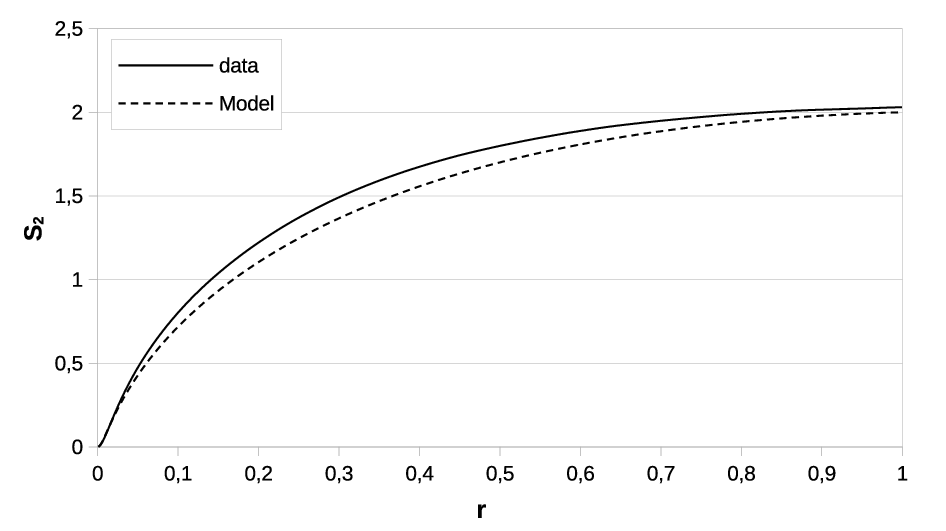}\hfill
		\includegraphics[width=0.5\textwidth]{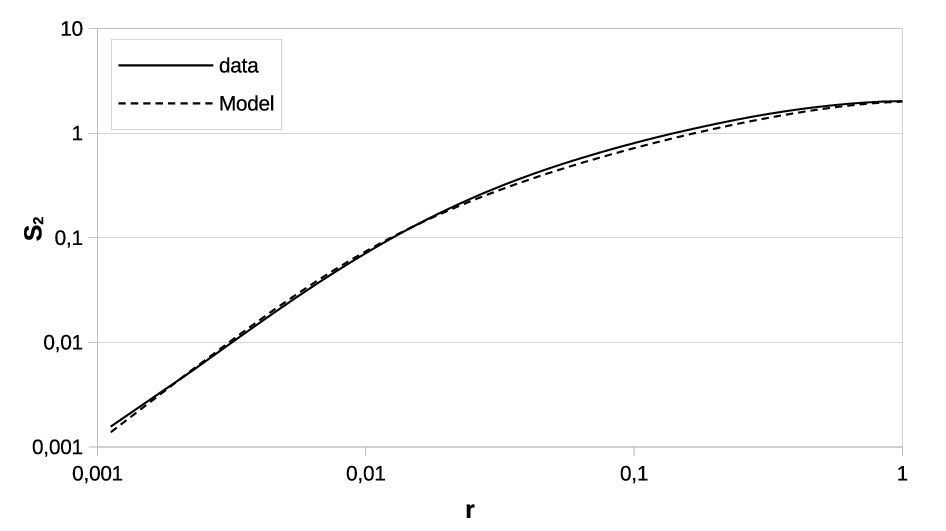}
	\caption{Second order structure function (straight lines) for the measurement with $Re=7\cdot 10^3$ in linear (left) and logarithmic (right) scale in comparison to the model prediction (dashed line) according to eq. (\ref{fullScaleS2}).}
	\label{s2forLowRe}
\end{figure}

The violation of the scaling law (\ref{rnDef}) for low Reynolds numbers as apparent in figure \ref{residualsForOtherRe} is mirrored in the model fits of structure functions for the two data sets. Figures \ref{s2forMediumRe} and \ref{s2forLowRe} show the second order structure functions in comparison to the model prediction (\ref{fullScaleS2}) for the data sets with and $Re \approx 2\cdot 10^4$ and $Re \approx 7\cdot 10^3$. Only the set with the higher Reynolds number of $2\cdot 10^4$ is in reasonable approximation described by the model, whereas the model fit for lower $Re$ exhibits significant deviations.

\section{Simplified and precise models for the second order structure function} \label{fullPrescisionS2}

The derivation of the closed full--scale model (\ref{closedFullScaleStwo}) made use of simplifying approximations for the function $d(r)$ and the third order structure function. These approximations yield a tractable equation for $S_2(r)$ but in turn introduce inaccuracies into the model. These approximations were introduced for convenience and can be replaced by more precise assumptions. This comes at the cost of analytical tractability but allows to quantify the error caused by these simplifications.

Equation (\ref{fullS2Eq}) for the second order structure function
\begin{eqnarray}
	\frac{\partial}{ \partial r} S_{2}(r) \; = \;  - \, \frac{Re}{6}\, \frac{\rho^2}{r^2} \, S_{3}(r) \, +  \, \frac{\zeta_{2}}{r} \, S_{2}(r) \, - \, 2 \, \zeta_2 \, d(r), \label{fullS2EqAgain}
\end{eqnarray}
needs to be completed by assumptions for $d(r)$ and an expression for $S_3(r)$.

The model for $d(r)$ proposed in the main body of this paper, eq. (\ref{dDef}), is incorrect in that it does not fulfill the large--scale condition $d(r=1)=1$. This can be rectified by introducing a normalization factor:
\begin{eqnarray}
	d(r) \; = \; r \, \sqrt{ \frac{1 + \rho^2}{r^2 + \rho^2} }. \label{fullDdef}
\end{eqnarray}

Closure of the model was achieved with the help of the four--fifth law (\ref{fourfifthDimensionless}). It is correct for the dissipative and inertial range, but obviously violates the large--scale condition $S_3(r=1)=0$. The issue can be fixed by combining the four--fifth law with the large--scale model (\ref{ExtS3Solution}) for $S_3(r)$:
\begin{eqnarray}
	S_3(r) \; & = & \; - \, \frac{4}{5} \, \epsilon \, r \left\{ \, 1 - \, \frac{F(r)}{F(1)} \right\} \; + \; \frac{6}{Re} \, \frac{\partial}{\partial r} \, S_2(r). \label{fourfifthExtended} 
\end{eqnarray}
with $F(r)$ as defined in (\ref{fOfRdef}).

\begin{figure}[ht]
	\centering
		\includegraphics[width=0.5\textwidth]{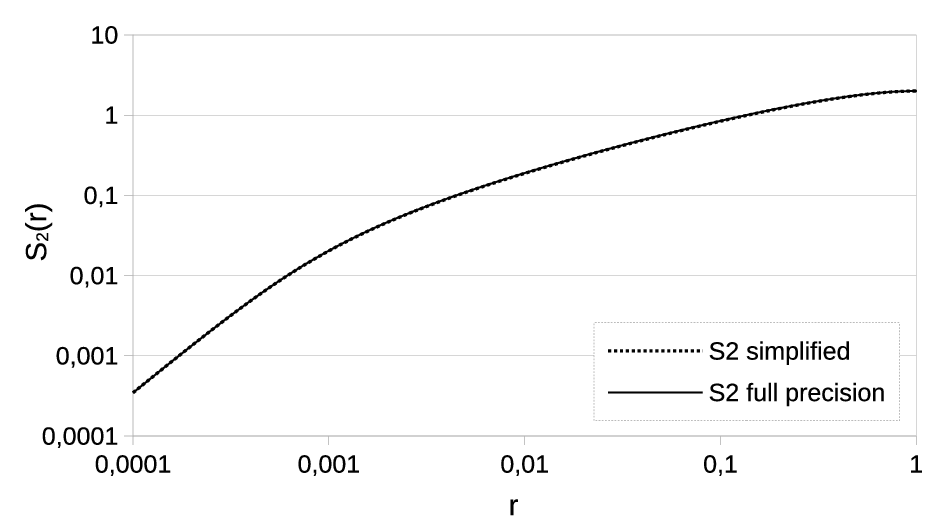}\hfill
		\includegraphics[width=0.5\textwidth]{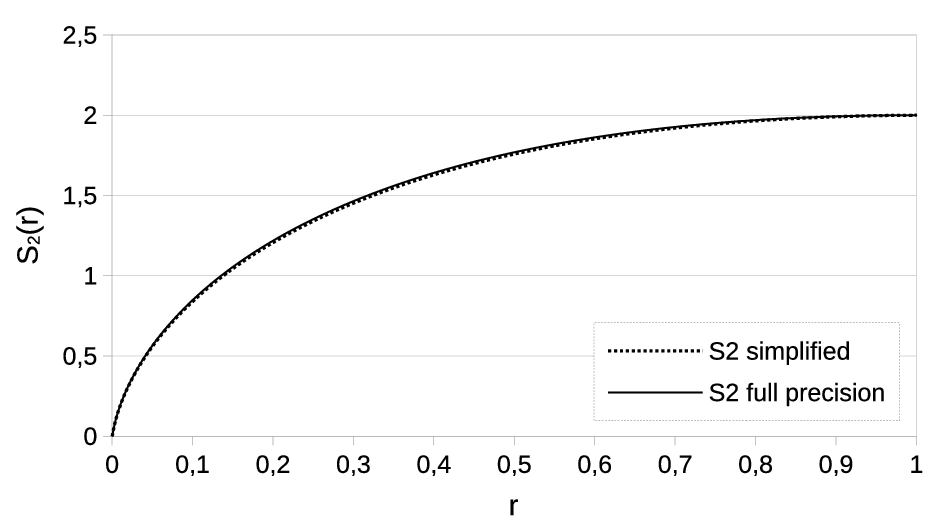}
	\caption{Second--order structure function in the simplified (dotted line) and full precision (straight line) model in linear (left) and logarithmic scale (right)}. 
	\label{fullVsSimplifiedS2Model}
\end{figure}

Equations (\ref{fullS2EqAgain})-(\ref{fourfifthExtended}) yield a closed full--scale model for $S_2(r)$ consistent with the large--scale boundary condition of a vanishing slope at the system length scale. A comparison of the numerical solution of this extended model with the simplified model according to eq. (\ref{closedFullScaleStwo}) is shown in figure \ref{fullVsSimplifiedS2Model}. When the scale parameter $\rho$ is optimized individually for the models to ensure that $S_2(1)=2$ in each case, we find good agreement between the two models with a maximum relative deviation of $1\%$. The optimized scale parameters are $\rho = 5.74 \cdot 10^{-4}$ and $\rho = 5.78 \cdot 10^{-4}$, respectively, corresponding to a relative deviation of only $0.7\%$. 

The method can also be used to examine the influence of the specific choice of sigmoid for $d(r)$. Replacing the algebraic function in (\ref{fullDdef}) with the hyperbolic tangent of $r / \rho$ leads to deviations that are of similar order of magnitude: The optimized values for $\rho$ differ by $0.5\%$, and the second order structure functions exhibit maximum relative deviations of $1\%$.

\bibliographystyle{ieeetr}
\bibliography{References}

\end{document}